\documentclass[aps,twocolumn,floats,superscriptaddress,prd,nofootinbib]{revtex4}
\usepackage{graphicx, epsfig, bm, amsmath}

\usepackage{color}
\usepackage{hyperref}
\usepackage{ifthen}
\usepackage{xstring}

\begin{document}

%%%%%%%%%%%%%%%%%%%%%%%%%%%%%%%%%%

\newcommand{\zmin}{z_{\rm min}}
\newcommand{\zmax}{z_{\rm max}}
\newcommand{\dz}{\Delta z}
\newcommand{\dzsub}{\Delta z_{\rm sub}}
\newcommand{\zminsn}{z_{\rm min}^{\rm SN}}
\newcommand{\nzpc}{N_{z,{\rm PC}}}
\newcommand{\nmax}{N_{\rm max}}
\newcommand{\amax}{a_{\rm max}}
\newcommand{\atr}{a_{\rm tr}}
\newcommand{\aeq}{a_{\rm eq}}

\newcommand{\wmin}{w_{\rm min}}
\newcommand{\wmax}{w_{\rm max}}
\newcommand{\wfid}{w_{\rm fid}}
\newcommand{\wwa}{w_0\text{--}w_a}

\newcommand{\lcdm}{$\Lambda$CDM}
\newcommand{\fom}{{\rm FoM}}

\newcommand{\thetal}{\bm{\theta}_{\Lambda}}
\newcommand{\thetaq}{\bm{\theta}_{\rm Q}}
\newcommand{\thetade}{\bm{\theta}_{\rm DE}}
\newcommand{\thetas}{\bm{\theta}_{\rm S}}
\newcommand{\thetadel}{\bm{\theta}_{{\rm DE},\Lambda}}
\newcommand{\thetadeq}{\bm{\theta}_{\rm DE,Q}}
\newcommand{\thetades}{\bm{\theta}_{\rm DE,S}}
\newcommand{\thetaother}{\bm{\theta}_{\rm nuis}}

\newcommand{\om}{\Omega_{\rm m}}
\newcommand{\ode}{\Omega_{\rm DE}}
\newcommand{\ok}{\Omega_{\rm K}}
\newcommand{\omhh}{\Omega_{\rm m} h^2}
\newcommand{\obhh}{\Omega_{\rm b} h^2}
\newcommand{\winf}{w_{\infty}}
\newcommand{\scrm}{\mathcal{M}}
\newcommand{\osf}{\Omega_{\rm sf}}
\newcommand{\omf}{\Omega_{\rm m}^{\rm fid}}
\newcommand{\scrmf}{\mathcal{M}^{\rm fid}}
\newcommand{\rhode}{\rho_{\rm DE}}
\newcommand{\rhoc}{\rho_{{\rm cr},0}}
\newcommand{\dlnl}{-2\Delta\ln\mathcal{L}}
\newcommand{\dlum}{d_{\rm L}}
\newcommand{\dlss}{D(z_*)}

\newcommand{\zh}{z_h}
\newcommand{\zbao}{z_{\rm BAO}}

\newcommand{\gpr}{G^{\prime}}
\definecolor{darkgreen}{cmyk}{0.85,0.2,1.00,0.2}

\newcommand{\mmcomment}[1]{\textcolor{red}{[{\bf MM}: #1]}}
\newcommand{\dhcomment}[1]{\textcolor{darkgreen}{[{\bf DH}: #1]}}
\newcommand{\wh}[1]{\textcolor{blue}{[{\bf WH}: #1]}}

%%%%%%%%%%%%%%%%%%%%%%%%%%%%%%%%%%%%%%%%%%%

\pagestyle{plain}

\title{Figures of merit for present and future dark energy probes}

\author{Michael J.\ Mortonson}
\affiliation{Center for Cosmology and AstroParticle Physics, 
        The Ohio State University, Columbus, OH 43210}

\author{Dragan Huterer}
\affiliation{Department of Physics, University of Michigan,
        Ann Arbor, MI 48109-1040}

\author{Wayne Hu}
\affiliation{Kavli Institute for Cosmological Physics, Department of Astronomy \& Astrophysics,
        and  Enrico Fermi Institute,
        University of Chicago, Chicago, IL 60637}
             
\begin{abstract}
We compare current and forecasted constraints on dynamical dark energy models
from Type Ia supernovae and the cosmic microwave background using figures of
merit based on the volume of the allowed dark energy parameter space.  For a
two-parameter dark energy equation of state that varies linearly with the
scale factor, and assuming a flat universe, the area of the error ellipse can
be reduced by a factor of $\sim 10$ relative to current constraints by future
space-based supernova data and CMB measurements from the Planck satellite. If
the dark energy equation of state is described by a more general basis of
principal components, the expected improvement in volume-based figures of
merit is much greater.  While the forecasted precision for any single
parameter is only a factor of 2--5 smaller than current uncertainties, the
constraints on dark energy models bounded by $-1\leq w\leq 1$ improve for
approximately 6 independent dark energy parameters resulting in a reduction of
the total allowed volume of principal component parameter space by a factor of
$\sim 100$.  Typical quintessence models can be adequately described by just
2--3 of these parameters even given the precision of future data, 
leading to a more modest but
still significant improvement.  In addition to advances in supernova and CMB
data, percent-level measurement of absolute distance and/or the expansion rate
is required to ensure that dark energy constraints remain robust to variations
in spatial curvature.
\end{abstract}

\maketitle

%=================================================================
\section{Introduction}
\label{sec:intro}

In the absence of physically compelling models for dark energy, figures of
merit (FoMs) are a useful tool that encapsulate the constraining power of
cosmological data.  FoMs combine various constraints on the expansion history
of the universe (e.g.\ the distance-redshift relation) in a single number, or
at most a handful of numbers, which serve as simple and quantifiable metrics
by which to evaluate current and proposed experiments
\cite{DETF,Albrecht_Bernstein,FoMSWG}.  The simplest schemes adopt fixed
functional forms for the evolution of the dark energy equation of state and
define the FoM as the inverse of the allowed parameter volume
\cite{Huterer_Turner}.  One widely used version is the two dimensional
$w_0$--$w_a$ parametrization \cite{DETF} but other higher dimensional
versions have also been considered
\cite{Huterer_Turner,Wang_FoM,Crittenden_Pogosian_Zhao}.

Any simple parametrization of the expansion history risks
biasing the FoM in favor of or against certain types of data 
by choosing a fixed functional form \cite{Albrecht_Bernstein}.
To avoid this problem, one can use
more complicated schemes that parametrize the whole functional freedom in the
dark energy equation of state evolution and separate the expansion history
and growth of structure information.  For example, uncorrelated modes of
piecewise-constant discretizations of the equation of state that are local in
redshift
\cite{Hu_PC,Huterer_Cooray,Wang_Tegmark_2005,Shapiro_Turner,Dick,Suletal07,Zhao_Huterer_Zhang,Zhao_Zhang:2009,Serra:2009}
or constructed from principal components (PCs)
\cite{Hu_PC,Huterer_Starkman,PaperI,PaperII,Huterer_Peiris,Tang:2008hm,Kitching:2009yr}
have been employed to characterize both current and future data.  In
particular, the inverse parameter volume of the PC amplitudes, defined
separately for each experiment, has been advocated as a FoM \cite{FoMSWG}.

Besides avoiding biasing results towards a particular functional form, 
a more model-independent FoM has the advantage of being able to identify 
improvements in dark energy constraints that might be missed by FoMs 
with fewer parameters.
On the other hand, not all improvements in a multidimensional PC FoM 
reflect improvements in constraining the space of known dark
energy models \cite{dePutter_Linder2,Barnard:2008mn}. 
For example, stronger constraints may exclude
regions of the parameter space not occupied by typical models.

In this paper, we study the FoMs defined both with the commonly-used
$w_0$--$w_a$ parametrization and with PCs based on forecasts for Type Ia
supernova (SN) and cosmic microwave background (CMB) data.  
Previous studies of PC-based FoMs have generally relied on the
Fisher matrix approximation, and the implementation and utility 
of these FoMs for real data has not been addressed.
Here we define straightforward generalizations of PC FoMs and apply them to
both actual data from current measurements and forecasts for future data.
While we still construct the PCs
using the Fisher matrix approach, we then treat the PC amplitudes as free
parameters and constrain them using full Markov Chain Monte
Carlo (MCMC) likelihood explorations with current or future data. This
methodology follows that employed in previous papers where we studied
generalized predictions of classes of dark energy models based on forecasts
\cite{PaperI} and current data \cite{PaperII}.  The new element here is the
application of these methods to the study of dark energy FoMs.

The standard approach for constructing the principal component-based FoM is to
use the PCs specific to the experiment and the cosmological probe(s)
considered; see e.g.~Ref.~\cite{FoMSWG}. However, this approach makes it
difficult to directly compare PC-based FoMs for different experiments and to
assess improvements in specific regions of parameter space.  To facilitate
such direct comparison between FoMs for current and future data, we choose
instead to fix the set of PCs based on a specific projection of future data
rather than computing separate PCs for future and current data.

This paper is organized as follows. In Sec.~\ref{sec:methods}, we describe the
current and forecasted SN and CMB data sets as well as additional priors from
baryon acoustic oscillations (BAO) and measurements of the Hubble constant.
We also briefly review the MCMC analysis methods used to infer dark energy
constraints. In Sec.~\ref{sec:results}, we compare FoMs from the current and
future SN and CMB constraints for both the commonly-used $w_0$--$w_a$ model
(Sec.~\ref{sec:fomw0wa}) and a more general PC-based parametrization
(Sec.~\ref{sec:fompc}). We summarize and discuss these results in
Sec.~\ref{sec:discussion}.

%=================================================================
\section{Methodology}
\label{sec:methods}

In this section, we review the current data sets and assumptions about future
experiments for forecasts that we use in this paper.  We refer the reader to
Refs.~\cite{PaperI} (hereafter MHH1) and~\cite{PaperII} (MHH2) for more
details concerning the forecasts and current data sets, respectively.  All
forecasts in this paper assume that the data originates from a flat
 cosmological constant ($w=-1$) model with present matter fraction
$\om=0.24$ and Hubble constant $H_0 = 73~{\rm km~s}^{-1}~{\rm Mpc}^{-1}$.

% ****************************************
\begin{figure}[t]
\centerline{\psfig{file=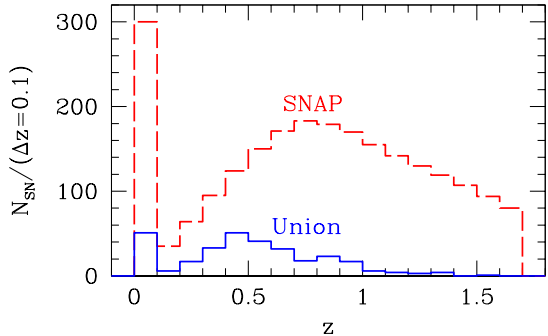, width=3.in}}
\caption{
Redshift distributions of Type Ia supernovae in the Union compilation 
(solid blue) and assumed for SNAP (dashed red), including the anticipated low-redshift 
sample from other surveys at $z<0.1$.
}
\label{fig:sndist}
\end{figure}
% ****************************************

%=================================================================
\subsection{Current SN and CMB Data}
\label{sec:current}

The Type Ia SN sample we use is the Union compilation~\cite{SCP_Union}.  These
SN observations measure relative distances, $D(z_1)/D(z_2)$, over a range of
redshifts spanning $0.015 \leq z \leq 1.551$, with most SNe at $z \lesssim 1$
(see Fig.~\ref{fig:sndist}).  We include SN constraints using the likelihood
code for the Union data sets \cite{Union_like}, which includes estimated
systematic errors in the covariance matrix \cite{SCP_Union}.

For the current CMB data, we use the 5-year data release from the WMAP
satellite \cite{Komatsu_2008,Nolta_2008,Dunkley_2008} employing the likelihood
code available at the LAMBDA web site \cite{WMAP_like}.  Unlike the CMB priors
used for the forecasts below, the likelihood used here contains the full
information from the CMB angular power spectra, except for the small effects
of gravitational lensing of the CMB that add little to current dark energy
constraints from WMAP.  We compute the CMB angular power spectra using the
code CAMB \cite{Lewis:1999bs,camb_url} modified with the parametrized
post-Friedmann (PPF) dark energy module \cite{PPF,ppf_url} to include models
with general dark energy equation of state evolution where $w(z)$ may cross
$w=-1$.

% ****************************************
\begin{figure}[t]
\centerline{\psfig{file=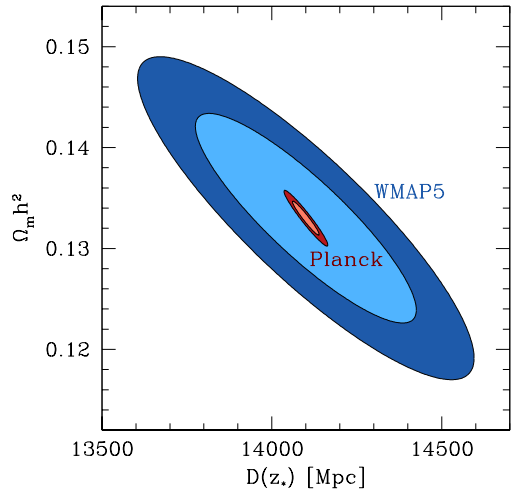, width=2.8in}}
\caption{
Approximate constraints on $D(z_*)$ and $\omhh$ from WMAP 5-year data (outer contours, blue shading) and 
forecasts for Planck (inner contours, red shading), showing 68\% CL (light shading) and 
95\% CL (dark shading) contours.
}
\label{fig:cmb}
\end{figure}
% ****************************************

%=================================================================
\subsection{SN and CMB Forecasts}
\label{sec:forecasts}

For our SN forecasts, we take the expected redshift distribution for the
SuperNova/Acceleration Probe (SNAP) \cite{KLMM_SNAP,SNAP} plus a low-$z$
sample of 300 SNe at $0.03<z<0.1$.  The SNAP magnitude errors include both
statistical and systematic components:
\begin{equation}
\sigma_\alpha^2 = \left(\frac{\dz}{\dzsub}\right)\left[\frac{0.15^2}{N_\alpha} +
0.02^2 \left(\frac{1+z_{\alpha}}{2.7}\right)^2 \right],
\label{eq:magerr}
\end{equation}
where $N_\alpha$, shown in Fig.~\ref{fig:sndist}, is the number of SNe in each
bin of width $\dz$ ($\dz=0.1$ except for the statistical uncertainties in the
low-$z$ SN bin, for which $\dz=0.1-\zminsn=0.07$), and $\dzsub$ is the width
of the sub-bins used to smooth the distribution of SNe in redshift. We use 500
sub-bins up to $\zmax=1.7$.  The second term on the right hand side of
Eq.~(\ref{eq:magerr}) models a systematic floor that increases linearly with
$z$ up to a maximum at $\zmax$ of $0.02$ mag per $\dz=0.1$ bin
\cite{LinHut_highz}.

For the Planck CMB forecasts, we use a $2\times 2$ covariance matrix
corresponding to the inner error ellipses in Fig.~\ref{fig:cmb},
\begin{equation}
{\bf C}^{\rm CMB} = 
\left( 
\begin{array}{cc}
(0.0018)^2 &  -(0.0014)^2 \\
-(0.0014)^2 & (0.0011)^2
\end{array}
\right),
\end{equation}
 with parameters 
\begin{equation}
{\bm \theta}^{\rm CMB} = \{\ln (\dlss/{\rm Mpc}),\,\omhh \}\,.
\label{eq:tildetheta}
\end{equation}
Here $\dlss$ is the comoving angular diameter distance to recombination.

In the Planck forecasts we ignore additional CMB information about dark energy
such as the ISW effect and gravitational lensing of the CMB.  Planck data are
expected to obtain limits on the fraction of early dark energy of
$\sigma(\ode(z_*))\approx 0.004$ \cite{dePutter:2009kn,Hollenstein:2009ph};
however, since these limits may depend on the modeling of early dark energy,
we include in our forecasts only a conservative prior of
$\sigma(\ode(z_*))=0.025$ (with fiducial value $\ode(z_*)\sim 10^{-9}$) which
approximates the current constraint from WMAP \cite{Doran}.

%=================================================================
\subsection{Additional Priors}
\label{sec:priors}

For both the current and forecasted constraints we add additional priors from
recent measurements of baryon acoustic oscillations and the Hubble constant.
The BAO constraint we use is based on the measurement of the correlation
function of SDSS Luminous Red Galaxies (LRGs) \cite{Eisenstein}, which
determines the distance and expansion rate at $\zbao\approx 0.35$ through the
combination $D_V(z) \equiv [z D^2(z)/H(z)]^{1/3}$.  We implement this
constraint by taking the volume average of this quantity, $\langle D_V \rangle$, over
the LRG redshifts, $0.16<z<0.47$, and comparing with the value of $A \equiv
\langle D_V \rangle \sqrt{\omhh}/\zbao$ given in Ref.~\cite{Eisenstein}, $A =
0.472 \pm 0.017$ (taking the scalar spectral tilt to be 
$n_s=0.96$). Using more recent BAO constraints, e.g.\ from 
Ref.~\cite{Percival09}, has only a small effect on the current constraints 
with SN and CMB data \cite{PaperII}.

We include the recent Hubble constant ($H_0$) constraint from the SHOES team
\cite{SHOES}, based on SN distances at $0.023<z<0.1$ that are linked to a
maser-determined absolute distance using Cepheids observed in both the maser
galaxy and nearby galaxies hosting Type Ia SNe.  The SHOES measurement
determines the absolute distance to a mean SN redshift of $\zh=0.04$,
which effectively corresponds to a constraint on $H_0$ for models with
relatively smooth dark energy evolution in the recent past (cf.~\cite{HubTrans})
 such that 
$\lim_{z\to 0} D(z) = cz/H_0$.  
We implement this constraint as a 
measurement of $D(\zh) = c\zh/(74.2 \pm 3.6$ km~s$^{-1}$~Mpc$^{-1}$).

%=================================================================
\subsection{MCMC Methodology}
\label{sec:mcmc}

Given the current or forecasted data, we use MCMC likelihood analysis
(e.g.\ see~\cite{Christensen:2001gj,Kosowsky:2002zt,Dunetal05}) to determine
dark energy parameter constraints and figures of merit for both the simple
$w_0$--$w_a$ models and general PC parametrization.  From the likelihood
${\cal L}({\bf x}|\bm{\theta})$ of the data ${\bf x}$ given each proposed
parameter set $\bm{\theta}$, Bayes' Theorem tells us the posterior probability
distribution of the parameter set given the data
\begin{equation}
{\cal P}(\bm{\theta}|{\bf x})=
\frac{{\cal L}({\bf x}|\bm{\theta}){\cal P}(\bm{\theta})}{\int d\bm{\theta}~
{\cal L}({\bf x}|\bm{\theta}){\cal P}(\bm{\theta})},
\label{eq:bayes}
\end{equation}
where ${\cal P}(\bm{\theta})$ is the prior probability density. The MCMC
algorithm generates random draws from the posterior distribution.  We test
convergence of the samples to a stationary distribution that approximates
${\cal P}(\bm{\theta}|{\bf x})$ by applying a conservative Gelman-Rubin
criterion \cite{gelman/rubin} of $R-1\lesssim 0.01$ across a minimum of four
chains for each model class.  We use the code CosmoMC
\cite{Lewis:2002ah,cosmomc_url} for the analysis of current data and an
independent MCMC code for forecasts.

%=================================================================
\section{Figure of Merit Comparisons}
\label{sec:results}

%=================================================================
\subsection{$w_0$--$w_a$ Figure of Merit}
\label{sec:fomw0wa}

\begin{figure}[t]
\centerline{\psfig{file=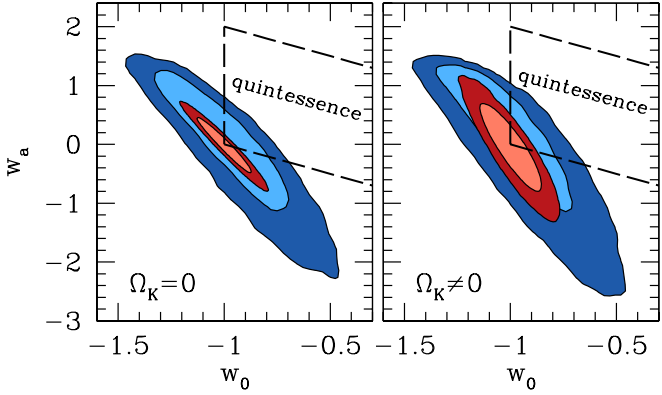, width=3.5in}}
\caption{ Constraints on $w_0$ and $w_a$, assuming a flat universe (left panel) or
  marginalizing over spatial curvature (right panel).  68\% CL (light shading)
  and 95\% CL (dark shading) regions are shown for both current Union+WMAP
  constraints (outer contours, blue shading) and SNAP+Planck forecasts (inner
  contours, red shading).  Dashed lines mark the boundary of the quintessence
  prior, $-1\leq w(z) \leq 1$.
}
\label{fig:w0wacontours}
\end{figure}

We first consider the two-parameter model for the dark energy equation 
of state \cite{Linder_wa,Chevallier_Polarski}
\begin{equation}
w(z) = w_0 + w_a \frac{z}{1+z}.
\label{eq:w0wa}
\end{equation}
The FoM for this model defined by the 
Dark Energy Task Force \cite{DETF,Huterer_Turner} is the
inverse of the area of the 95\%~CL region $A_{95}$ in the $w_0$--$w_a$ plane.
Figure~\ref{fig:w0wacontours} shows these regions for the current and forecasted
data with and without marginalization of spatial curvature $\ok$.

For a Gaussian error distribution, $A_{95}$ is proportional to the square root
of the determinant of the 2D covariance matrix ${\bf C}$ for $w_0$ and $w_a$.
Since the constant of proportionality used in practice for the FoM varies
widely in the literature (e.g., see \cite{Bassett:2009uv}), we simply define
\begin{eqnarray}
\fom^{(\wwa)}& \equiv & (\det {\bf C})^{-1/2}\nonumber\\
& \approx &
 {6.17 \pi \over A_{95}} \label{eq:fomw0wa} .
\end{eqnarray}
The approximate equality between the two lines in Eq.~(\ref{eq:fomw0wa})
becomes exact for a Gaussian posterior distribution.  Although the posterior
in $w_0$ and $w_a$ is not perfectly Gaussian, the FoM computed
using $\det {\bf C}$ in Eq.~(\ref{eq:fomw0wa}) remains a good approximation to
the area-based FoM.  The difference between the two is $\sim 10\%$
in the worst case (current data with $\ok\ne 0$) and $\lesssim 2\%$ in the
other cases.  Values of the $\det {\bf C}$ version of $\fom^{(\wwa)}$ are
given in Table~\ref{tab:w0wa}.  The FoMs for current data are
consistent with those found in previous studies of $w_0$--$w_a$ constraints
from similar data sets (e.g., \cite{Davetal07,Wright2007,Wang_FoM,Mantz}).

% ****************************************
% ****************************************

% ****************************************
\begin{table}
\caption{Figures of merit for $w_0$--$w_a$ models and 10-PC quintessence models 
without early dark energy.}
\begin{center}
\begin{tabular*}{\columnwidth}{@{\extracolsep{\fill}}lrrrr}
\hline
\hline
 & \multicolumn{2}{c}{$\fom^{(\wwa)}$} & \multicolumn{2}{c}{$\fom^{({\rm PC})}_{10}$} \\
Data & $\ok=0$ & $\ok\ne 0$ & $\ok=0$ & $\ok\ne 0$ \\
\hline
SNAP+Planck & 160 & 46 & 53000 & 19000 \\
Union+WMAP & 15 & 11 & 370 & 260 \\
\hline
ratio & 11 & 4.3 & 140 & 73 \\
\hline
\hline
\end{tabular*}
\end{center}
\label{tab:w0wa}
\end{table}
% ****************************************

While analyses of $w_0$--$w_a$ models typically allow $w(z)$ to cross $-1$, it
is useful to also consider a more restricted class of models that satisfy the
quintessence bound $-1\leq w\leq 1$ in light of the PC description below.
Imposing this quintessence prior greatly restricts the allowed parameter space
as shown in Fig.~\ref{fig:w0wacontours}.  Assuming a flat (nonflat) universe,
the $\det {\bf C}$ FoM is a factor of $\sim 8$ (6.5) larger than without the
quintessence prior for current data, and a factor of $\sim 13$ (9) larger for
forecasts.  Thus the addition of the quintessence prior increases the ratio of
future to current $\fom$ values by about 60\% (35\%) relative to the ratios in
Table~\ref{tab:w0wa}.  Note that the effect of the quintessence prior on the
FoM for forecasts depends on the choice of the true model from which the data is drawn; for example, had we
chosen fiducial $w_0$ and $w_a$ values that lie as far within the quintessence
prior boundaries as the current data allow, the forecasted area allowed within the
priors would be greater and the improvement in the FoM relative to current
constraints would be smaller.

Even in this case where the posterior distribution with the quintessence prior
is far from Gaussian, $\det {\bf C}$ still approximates the area-based FoM
through Eq.~(\ref{eq:fomw0wa}) reasonably well.  Although values of the $\det
{\bf C}$ FoM are smaller by 35--40\% than the area FoM, the ratio of future to
current FoM values is nearly unchanged.  We shall see that this type of
agreement for quintessence models carries over to to the more general PC-based
FoM in the next section.

% ****************************************
\begin{figure}[b]
\centerline{\psfig{file=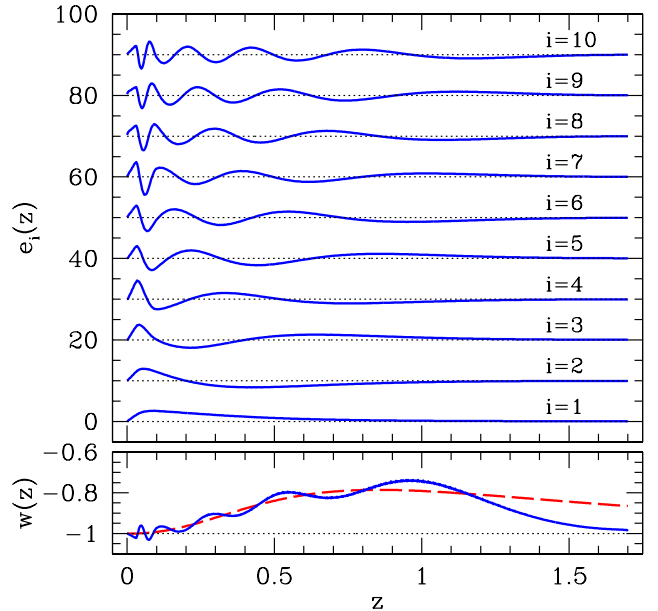, width=3.5in}}
\caption{Top panel: The first 10 PCs of $w(z)$ (increasing variance from
  bottom to top), with 500 redshift bins between $z=0$ and $\zmax=1.7$.  The
  PCs are offset vertically from each other for clarity.  Bottom panel: an
  example of $w(z)$ for a quintessence model (red dashed curve; see
  Eq.~(\ref{eq:qmodel})) and its representation using the first 10 PCs in
  Eq.~(\ref{eq:wpc}) (blue solid curve). Dotted lines show the $w=-1$ zero
  point for each component (top) and for the example model (bottom).  }
\label{fig:pcs}
\end{figure}
% ****************************************

% ****************************************
\begin{table}
\caption{Top hat prior rms $\Delta \alpha_i/\sqrt{12}$ and mean and 
rms of $\alpha_i$ from current data and forecasts, assuming 
flat quintessence models without early dark energy.}
\begin{center}
\begin{tabular*}{\columnwidth}{@{\extracolsep{\fill}}lrrrrr}
\hline
\hline
 & & \multicolumn{2}{c}{Union+WMAP} & \multicolumn{2}{c}{SNAP+Planck} \\
PC $i$ & $\Delta \alpha_i/\sqrt{12}$ & $\bar{\alpha}_i$ & $\sigma_i$ & $\bar{\alpha}_i$ & $\sigma_i$ \\
\hline
1 & 0.375 & \quad 0.061 & 0.041 & \quad 0.011 & 0.008 \\
2 & 0.421 & 0.087 & 0.132 & $-0.013$ & 0.037 \\
3 & 0.428 & 0.165 & 0.203 & 0.011 & 0.086 \\
4 & 0.411 & 0.206 & 0.278 & $-0.040$ & 0.141 \\
5 & 0.450 & 0.184 & 0.278 & $-0.028$ & 0.206 \\
6 & 0.452 & $-0.069$ & 0.394 & 0.053 & 0.284 \\
7 & 0.425 & $-0.071$ & 0.360 & 0.117 & 0.343 \\
8 & 0.454 & $-0.063$ & 0.436 & 0.043 & 0.374 \\
9 & 0.461 & $-0.281$ & 0.438 & $-0.147$ & 0.418 \\
10 & 0.463 & 0.026 & 0.448 & $-0.003$ & 0.424 \\
\hline
\hline
\end{tabular*}
\end{center}
\label{tab:pc}
\end{table}
% ****************************************

% ****************************************
\begin{figure*}[t]
\centerline{\psfig{file=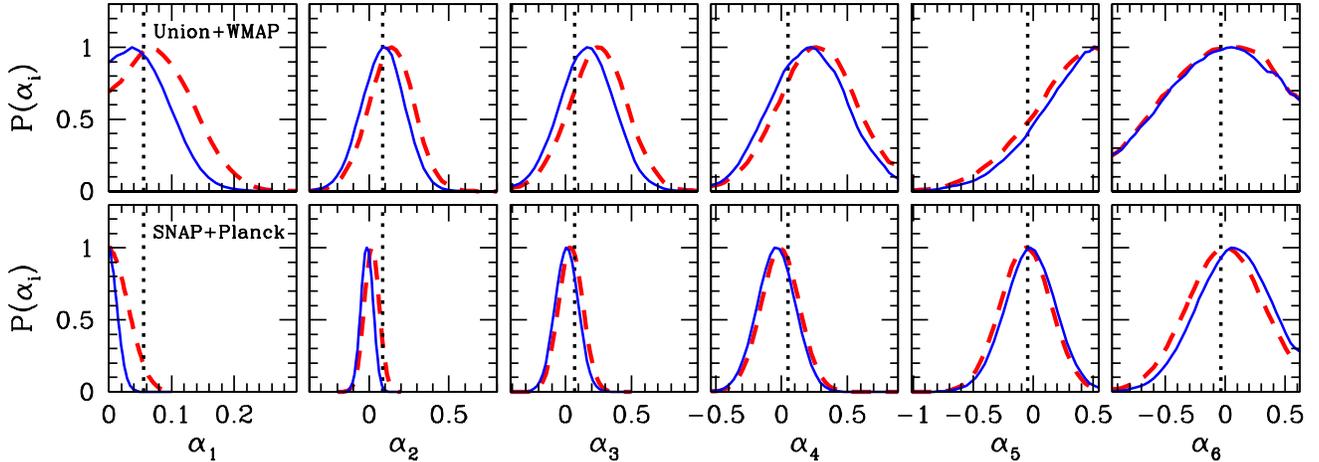, width=7in}}
\caption{Marginalized 1D posterior distributions for the first 6 PCs of flat
  (solid blue curves) and nonflat (dashed red curves) quintessence models
  without early dark energy. Marginalizing over the early dark energy
  parameter $\winf$ has little effect on the distributions. Top row: current
  Union+WMAP data; bottom row: forecasts for SNAP+Planck assuming a 
  realization of the data with $\alpha_i=0$.
  Plot boundaries that cut off the
  distributions at nonzero probability correspond to the top hat priors on
  $\alpha_i$ for quintessence models.  Higher variance PCs are mainly limited
  by the quintessence priors for both current constraints and forecasts (see
  Table~\ref{tab:pc}).  Vertical dotted lines show the predictions of an
  example quintessence model from Eq.~(\ref{eq:qmodel}) (see
  Fig.~\ref{fig:pcs}).  }
\label{fig:pc1d}
\end{figure*}
% ****************************************

%=================================================================
\subsection{Principal Component Figure of Merit}
\label{sec:fompc}

We generalize the dark energy parametrization to allow arbitrary variations of
the equation of state at $z<\zmax$ with a basis of principal components (PCs)
\cite{Hu_PC,Huterer_Starkman}.  Details of the PC construction can be found in
MHH1; here we highlight the points of special relevance for the FoM.

We construct the PCs based on the SN and CMB forecasts from
Sec.~\ref{sec:forecasts}. The PCs are a set of orthogonal functions ordered by
the precision with which they can be measured by the future SN and CMB data.
Specifically, the principal component functions $e_i(z_j)$ are eigenvectors of
the SNAP+Planck covariance matrix in the space of piecewise constant values of
the equation of state in redshift bins, $w(z_j)$. The principal components
form a basis in which an arbitrary function $w(z_j)$ may be expressed as
\begin{equation}
w(z_j) = -1 + \sum_{i=1}^{\nzpc} \alpha_i e_i(z_j),
\label{eq:wpc}
\end{equation}
where $\alpha_i$ are the PC amplitudes, $\nzpc = 1+\zmax/\dz$ is the number of
redshift bins of width $\dz$, and $z_j = (j-1) \dz$. We choose the maximum
redshift for variations in $w(z)$ to be $\zmax=1.7$, matching the largest
redshift for the SNAP supernova data. The impact of dark energy evolution at
higher redshifts is expected to be small, but perhaps non-negligible; to
account for this possibility, we parametrize the equation of state at
$z>\zmax$ by a constant, $\winf$.  Likewise, we consider models with spatial
curvature, $\ok\ne 0$. Note that $\winf$ and $\ok$ are not
allowed to vary from their fiducial values of $-1$ and $0$,
respectively, in the Fisher analysis used to construct the PCs.

% ****************************************
\begin{figure}[t]
\centerline{\psfig{file=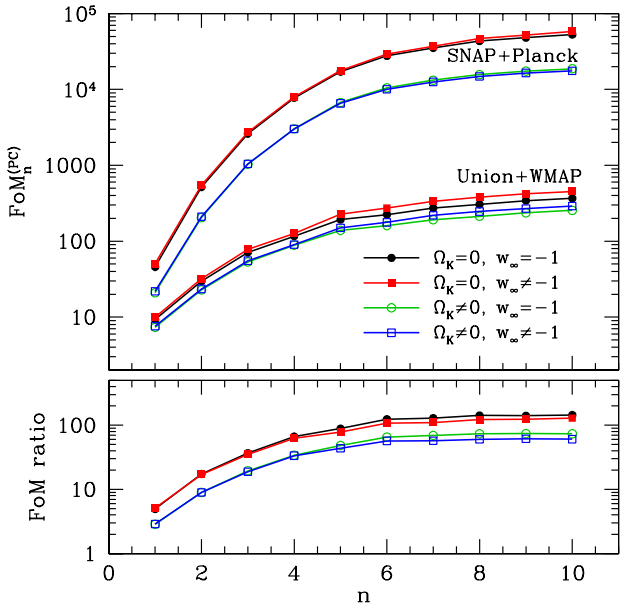, width=3.5in}}
\caption{ Top panel: PC figures of merit $\fom_n^{({\rm PC})}$ with forecasted
  uncertainties for SNAP+Planck and with measured uncertainties for
  Union+WMAP, normalized as in Eq.~(\ref{eq:fompc1}) to account for the
  quintessence prior.  Bottom panel: Ratios of $\fom_n^{({\rm PC})}$ forecasts
  to current values.  In both panels, point types indicate different
  quintessence model classes: flat (solid points) or non-flat (open points),
  either with (squares) or without (circles) early dark energy.  }
\label{fig:fompcratio}
\end{figure}
% ****************************************

Since the highest-variance PCs correspond to modes of $w(z)$ to which even future
data are insensitive, we truncate the sum in Eq.~(\ref{eq:wpc}) by 
replacing $\nzpc$ with
$\nmax<\nzpc$.  As shown in MHH1, the 10 lowest-variance PCs ($\nmax=10$) form
a basis which, for the classes of models we consider here, is sufficiently
complete for the forecasts. Therefore, 10 PCs more than suffice for the
current data as well.  This set of basis functions is displayed in Fig.~\ref{fig:pcs}.

We impose priors on the PC amplitudes corresponding to the range of $w(z)$
allowed for scalar field quintessence, $-1\leq w\leq 1$ following MHH1.  Our
conservative implementation excludes only parameter values that must violate
these bounds even when considering possible compensation from the omitted
higher-variance PCs (e.g.\ see the lower panel of Fig.~\ref{fig:pcs} at
$z<0.1$). This approach yields top hat priors of width
\begin{equation}
\Delta \alpha_i = {2 \over N_{z,\rm PC} }\sum_{j=1}^{N_{z,\rm PC}}  | e_i(z_j)|\,,
\end{equation}
which follows from Eq.~(A10) in MHH1. 

In analogy to the $w_0$--$w_a$ FoM in the previous section, we base the FoM
for dark energy PCs on the determinant of the covariance matrix of $\alpha_i$
for the $n$ lowest-variance PCs, $\det {\bf C}_n$.  
Even without informative data, this determinant is finite:
\begin{equation}
\det {\bf C}_n^{({\rm prior})} =  \prod_{i=1}^n\left(  {\Delta \alpha_i \over \sqrt{12}} \right)^{2} ,
\end{equation}
where the individual factors of $\Delta\alpha_i/\sqrt{12}$ (listed in
Table~\ref{tab:pc}) are the rms values of the corresponding top hat priors.
These priors impact the FoM in a manner similar to those imposed in
Ref.~\cite{FoMSWG}, where a Gaussian prior is adopted that requires the rms
variation of $1+w_i$ for each PC mode, averaged over scale factor, to be no
more than unity.  However, our priors are slightly stronger since we use the
quintessence bounds to impose a top hat prior on $1+w(z)$ at all redshifts.
In order that the FoM values do not reflect information that comes exclusively
from the quintessence prior, we follow the convention in Ref.~\cite{FoMSWG}
and renormalize the $\det {\bf C}$ statistic to obtain
\begin{equation}
\fom^{({\rm PC})}_n \equiv 
\left({ \det {\bf C}_n  \over \det {\bf C}_{n}^{(\rm prior)}}  \right)^{-1/2} \,.
\label{eq:fompc1}
\end{equation}

Table~\ref{tab:pc} shows the mean and rms of each of the 10 PCs from 
MCMC likelihood analysis using the
current data and forecasts.  Note that for components 6--10 in the current data
the rms is dominated by the prior, and likewise for 7--10 in the forecast.  
Figure~\ref{fig:pc1d} shows the one-dimensional posterior probability
distributions of the first 6 PCs where the information from the data resides.

Figure~\ref{fig:fompcratio} (upper panel) shows the FoM for the current data and
forecasts.  Note that for both data sets the FoM starts to saturate around the 6th
PC as expected.   The lower panel shows the ratio of FoMs of 
future and current data.  Once again the saturation point is around the 6th PC
with the total level of improvement varying from a factor of $\sim 60$ to $\sim 140$
depending mainly on whether spatial curvature is included.

This dependence of the FoM improvement on curvature is mainly due to a
degeneracy between $\ok$ and the first PC which reduces 
the FoM when marginalizing over $\ok$, particularly for the forecasts.  
This degeneracy is largely a consequence of our choice not to 
marginalize over $\ok$ when
constructing the PCs, but the leading degeneracy between dark
energy and curvature is included in the first PC.  
There is a related difference between the flat and
nonflat cases in the $w_0$--$w_a$ contours in Fig.~\ref{fig:w0wacontours}
and $\fom^{(w_0\text{--}w_a)}$ in Table~\ref{tab:w0wa}.

% ****************************************
\begin{figure}[t]
\centerline{\psfig{file=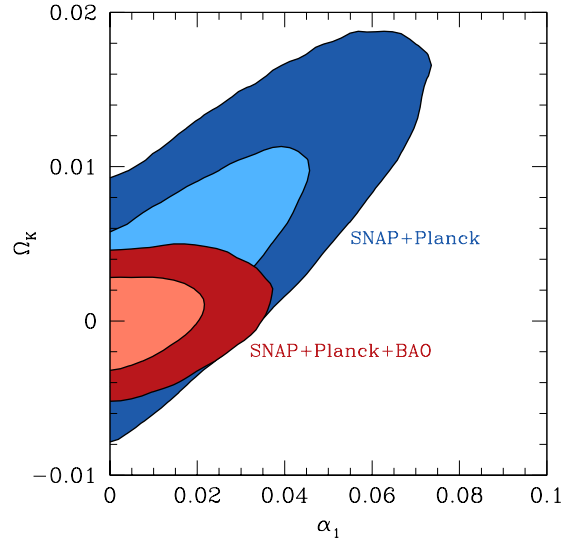, width=3.in}}
\caption{
Forecasts for $\ok$ and the amplitude of the first PC, $\alpha_1$, 
showing 68\%~CL (light shading) and 95\%~CL (dark shading) regions.
The large contours (blue shading) include only SNAP and Planck data
as well as the additional priors from Sec.~\ref{sec:priors}. 
The small contours (red shading) add to these data a 1\% measurement
of $\langle D_V \rangle$ averaged over the redshift bin $0.8<z<1.2$ 
as might be obtained from a future BAO 
experiment, reducing the degeneracy between curvature and dark energy.
}
\label{fig:omkvsa1}
\end{figure}
% ****************************************

Current BAO and $H_0$ measurements (Sec.~\ref{sec:priors}) constrain $\ok$ enough that current
SN+CMB PC uncertainties are affected little by curvature, but for the SN+CMB
forecasts we need $\sim 1\%$ measurements of an absolute distance scale 
to achieve a FoM improvement comparable to that in the flat case.
Figure~\ref{fig:omkvsa1} shows an example of breaking the $\ok$--$\alpha_1$ 
degeneracy using a 1\% $\langle D_V \rangle$
measurement at $z=0.8$--$1.2$ as might be achieved from a future BAO experiment.
As shown in Fig.~7 of MHH1, the shift in absolute distance corresponding
to the curvature degeneracy is largely independent of redshift and its
elimination could also be achieved with comparable measurements at $z=0-0.1$ from
improved Hubble constant probes or $z=3$ from high-redshift BAO.

The small gap in $\fom_n^{({\rm PC})}$ ratios in Fig.~\ref{fig:fompcratio} 
between models with and without early dark energy is 
driven by the current constraint on the 5th PC; as shown in 
Fig.~\ref{fig:pc1d}, $P(\alpha_5)$ is not centered on zero 
and the distribution is cut off by the prior. Including early dark energy
shifts this distribution further outside the prior, reducing the PC volume 
(and increasing the FoM) for current data. Forecasts for PCs are basically 
unaffected by early dark energy (see Fig.~\ref{fig:fompcratio}), 
so the overall effect of early dark energy is a slightly smaller 
FoM ratio when the 5th PC is included.
This particular feature is not generic in the sense that it would not 
necessarily show up for other choices of data sets or dark energy 
parametrizations, but similar effects could appear 
in other analyses where the dark energy priors play an important role.

Since we are using the PCs constructed assuming the forecasted data rather
than the current data, we could in principle have strong covariances between the
PC amplitudes that would be hidden in Fig.~\ref{fig:pc1d}.  Likewise, we could
have substantial differences between $\fom_n^{({\rm PC})}$ and definitions 
of the PC FoM involving ratios of $\sigma_i$
\cite{Albrecht_Bernstein,FoMSWG}.  
In practice, however, the covariances between PC amplitudes remain small 
in all cases. Even for current data, the difference between 
$(\det {\bf C}_n)^{-1/2}$ and $\prod_{i=1}^n \sigma_i^{-1}$ 
is $\lesssim 20\%$ for flat quintessence models and 
$\lesssim 30\%$ for nonflat models. The effect of covariances on the 
forecast FoM values is even smaller.
Thus the 1D distributions in Fig.~\ref{fig:pc1d} accurately depict the 
current and future constraints on PC amplitudes.

A separate question is whether the use of 
$\det {\bf C}$ gives misleading results due to non-Gaussianity of 
the PC posterior distributions (for example, due to distributions 
being cut off by the quintessence prior). To test the significance 
of such effects, we consider an alternate FoM 
analogous to the area-based FoM for $w_0$--$w_a$ models,
\begin{equation}
V_n^{-1} \equiv (2\sqrt{\pi})^n \int d{\bm{\alpha}}\, 
{\cal P}^2(\bm{\alpha}|{\bf x})\,,
\label{eq:fompc2}
\end{equation}
where $\bm{\alpha}$ is the parameter subset consisting of the 
first $n$ PC amplitudes.
The motivation of this form is that the allowed volume in parameter space 
is proportional to the inverse of the average number density
  of MCMC samples. The number density is proportional to ${\cal P}$, and the
  averaging over the posterior probability gives another factor of ${\cal
    P}$.  The normalization of Eq.~(\ref{eq:fompc2}) is chosen so that 
 for an $n$D Gaussian posterior with covariance ${\bf C}_n$,
    $V_n^{-1} = (\det {\bf C}_n)^{-1/2}$.

For the 4--5 lowest-variance PCs, we find good agreement between $V_n^{-1}$
and $(\det {\bf C}_n)^{-1/2}$ with differences of no more than 
$\sim 30\%$. Using the $V_n^{-1}$ FoM 
increases the ratio of future to current FoM values by 10--15\%.
For $n>5$, accurate computation of the integral in Eq.~(\ref{eq:fompc2})
becomes more difficult due to the sparsity of MCMC samples.

Improvements in $\fom_n^{({\rm PC})}$ do not necessarily represent 
significant improvements in the ability to
limit the parameter space of known dark energy models, especially for 
the higher 
PCs \cite{dePutter_Linder2,Barnard:2008mn}.   In order to address such issues,
one can project the predictions of a model for $w(z)$ onto the PC space and examine
whether the predictions lie in the volume excluded by forecasted constraints. 
As an illustrative example,
we consider a quintessence model with the potential 
\begin{equation}
V(\phi) = V_0 + {1\over 2}m^2 \phi^2.
\label{eq:qmodel}
\end{equation}  
This model provides examples in the thawing class \cite{Caldwell_Linder} for
$V_0\rightarrow 0$ and can also mimic the low redshift behavior of
Albrecht-Skordis models with oscillations around an offset minimum
\cite{AlbSko00}.  We consider an example where $m = 7 \times 10^{-33}$~eV and $V_0
/\rho_{\rm crit}= 0.717$ with $\Omega_{\rm DE}=0.733$, $h=0.69$ and
Hubble-drag frozen initial conditions $\dot \phi_i=0$; the equation of state
for this model is shown in the lower panel of Fig.~\ref{fig:pcs}.  These
parameters are chosen to be allowed by the current data but testable with
future data.

Figure~\ref{fig:pc1d} compares the predictions for this model with the data
constraints.  Note that even for the future data only the first two PCs are
stimulated at a level that the data can constrain.  Thus for these types of
models FoM$^{\rm (PC)}_2$ is more representative of the parameter volume
improvements than higher-dimensional FoMs.  
Taking into account the quintessence prior, the improvement
from current to future data in this case 
is comparable to that implied by the $w_0$--$w_a$ FoM.

More generally, studies have found that up to 3 PCs are useful for
distinguishing amongst different commonly-used quintessence models with data
sets comparable to our future forecasts \cite{Barnard:2008mn}.  On the other
hand, these studies do not preclude the possibility that FoM improvements in
the higher components can distinguish between other yet to be
investigated classes of models.

\section{Discussion}
\label{sec:discussion}

We have compared current and forecasted figures of merit (FoMs) for dark
energy using both a simple $w_0$--$w_a$ description of the equation of state,
and a more complicated but more complete principal component (PC)
parametrization.  By consistently using a fixed set of PCs based
on future data, and by generalizing the FoM definition to include possible
parameter covariance, we showed how PC FoMs can be applied to likelihood
analyses of both existing data sets and forecasts.

We have also shown that the covariance-based FoMs accurately represent
relative changes in the parameter volume, even in the presence of strongly
non-Gaussian posterior distributions such as those caused by
imposing top hat priors.  Traditional variance-based PC FoMs are
consistent with those that include the full covariance for the
cases we have tested, but we do not expect such agreement to hold in general.
For example, if our PC eigenfunctions from SN and CMB data are applied to
qualitatively different data, e.g.\ weak lensing or BAO, the FoMs
that account for the parameter covariances should be employed.

For the $w_0$--$w_a$ FoM, future space-based supernova data and CMB
measurements from the Planck satellite can improve on current measurements by
a factor of $\sim 10$.  For the PC FoM, the expected improvement is much
greater still.  While the forecasted precision for any single principal
component is only a factor of 2--5 smaller than current uncertainties, the
constraints on general quintessence models bounded by $-1\leq w\leq 1$ 
improve for approximately 6 components
resulting in a reduction of the total allowed volume of dark energy parameter
space by a factor of $\sim 60$--$140$. The FoM improvement depends mainly on
whether or not variations in spatial curvature are allowed, with the maximum
ratio of $\sim 140$ requiring either fixing the curvature with a theoretical
prior ($\ok=0$) or combining future SN and CMB data with a measurement of
absolute distance and/or the expansion rate with at least $\sim 1\%$ accuracy.

Although improvements from future SN and CMB data extend to 6 components of $w(z)$, 
many commonly-considered quintessence models are adequately
described by the first three PCs for which the allowed volume only decreases by
a factor of $\sim 20$--$40$ relative to current measurements.
While figures of merit provide a convenient model-independent metric to
evaluate experiments, they must be interpreted with caution when science goals
include discriminating amongst specific dark energy models.

\vspace{1cm} {\it Acknowledgments:} MJM was supported by CCAPP at Ohio State;
DH by the DOE OJI grant under contract DE-FG02-95ER40899, NSF under contract
AST-0807564, and NASA under contract NNX09AC89G; WH by the KICP under NSF
contract PHY-0114422, DOE contract DE-FG02-90ER-40560 and the Packard
Foundation.

\vfill
\bibliographystyle{arxiv_physrev}
\bibliography{pcfom}

\def\eprinttmppp@#1arXiv:@{#1}
\providecommand{\arxivlink[1]}{\href{http://arxiv.org/abs/#1}{arXiv:#1}}
\providecommand{\arxivlinknopre[1]}{\href{http://arxiv.org/abs/#1}{#1}}
\providecommand{\eprintmod}[1][XXXX.XXXX]{\IfSubStr{#1}{arXiv}{\arxivlinknopre%
{#1}}{\arxivlink{#1}}}
\providecommand{\adsurl}[1]{\href{#1}{ADS}}
\begin{thebibliography}{58}
\expandafter\ifx\csname natexlab\endcsname\relax\def\natexlab#1{#1}\fi
\expandafter\ifx\csname bibnamefont\endcsname\relax
  \def\bibnamefont#1{#1}\fi
\expandafter\ifx\csname bibfnamefont\endcsname\relax
  \def\bibfnamefont#1{#1}\fi
\expandafter\ifx\csname citenamefont\endcsname\relax
  \def\citenamefont#1{#1}\fi
\expandafter\ifx\csname url\endcsname\relax
  \def\url#1{\texttt{#1}}\fi
\expandafter\ifx\csname urlprefix\endcsname\relax\def\urlprefix{URL }\fi

\bibitem{DETF}
A.~Albrecht {\em et~al.},
\newblock \eprintmod[astro-ph/0609591].
%%CITATION = ASTRO-PH 0609591;%%

\bibitem{Albrecht_Bernstein}
A.~J. Albrecht and G.~Bernstein,
\newblock Phys. Rev. {\bf D75}, 103003 (2007), [\eprintmod[astro-ph/0608269]].
%%CITATION = ASTRO-PH/0608269;%%

\bibitem{FoMSWG}
A.~J. Albrecht {\em et~al.},
\newblock \eprintmod[0901.0721].
%%CITATION = 0901.0721;%%

\bibitem{Huterer_Turner}
D.~Huterer and M.~S. Turner,
\newblock Phys. Rev. {\bf D64}, 123527 (2001), [\eprintmod[astro-ph/0012510]].
%%CITATION = ASTRO-PH 0012510;%%

\bibitem{Wang_FoM}
Y.~Wang,
\newblock Phys. Rev. {\bf D77}, 123525 (2008), [\eprintmod[0803.4295]].
%%CITATION = 0803.4295;%%

\bibitem{Crittenden_Pogosian_Zhao}
R.~G. Crittenden, L.~Pogosian and G.-B. Zhao,
\newblock JCAP {\bf 0912}, 025 (2009), [\eprintmod[astro-ph/0510293]].
%%CITATION = ASTRO-PH/0510293;%%

\bibitem{Hu_PC}
W.~Hu,
\newblock Phys. Rev. {\bf D66}, 083515 (2002), [\eprintmod[astro-ph/0208093]].
%%CITATION = ASTRO-PH 0208093;%%

\bibitem{Huterer_Cooray}
D.~Huterer and A.~Cooray,
\newblock Phys. Rev. {\bf D71}, 023506 (2005), [\eprintmod[astro-ph/0404062]].
%%CITATION = ASTRO-PH 0404062;%%

\bibitem{Wang_Tegmark_2005}
Y.~Wang and M.~Tegmark,
\newblock Phys. Rev. {\bf D71}, 103513 (2005), [\eprintmod[astro-ph/0501351]].
%%CITATION = ASTRO-PH 0501351;%%

\bibitem{Shapiro_Turner}
C.~Shapiro and M.~S. Turner,
\newblock Astrophys. J. {\bf 649}, 563 (2006), [\eprintmod[astro-ph/0512586]].
%%CITATION = ASTRO-PH/0512586;%%

\bibitem{Dick}
J.~Dick, L.~Knox and M.~Chu,
\newblock JCAP {\bf 0607}, 001 (2006), [\eprintmod[astro-ph/0603247]].
%%CITATION = ASTRO-PH 0603247;%%

\bibitem{Suletal07}
D.~{Sarkar} {\em et~al.},
\newblock Phys. Rev. Lett. {\bf 100}, 241302 (2008), [\eprintmod[0709.1150]].

\bibitem{Zhao_Huterer_Zhang}
G.-B. Zhao, D.~Huterer and X.~Zhang,
\newblock Phys. Rev. {\bf D77}, 121302 (2008), [\eprintmod[0712.2277]].
%%CITATION = 0712.2277;%%

\bibitem{Zhao_Zhang:2009}
G.-B. Zhao and X.-m. Zhang,
\newblock Phys. Rev. {\bf D81}, 043518 (2010), [\eprintmod[0908.1568]].
%%CITATION = 0908.1568;%%

\bibitem{Serra:2009}
P.~Serra {\em et~al.},
\newblock Phys. Rev. {\bf D80}, 121302 (2009), [\eprintmod[0908.3186]].
%%CITATION = 0908.3186;%%

\bibitem{Huterer_Starkman}
D.~Huterer and G.~Starkman,
\newblock Phys. Rev. Lett. {\bf 90}, 031301 (2003),
  [\eprintmod[astro-ph/0207517]].
%%CITATION = ASTRO-PH 0207517;%%

\bibitem{PaperI}
M.~J. {Mortonson}, W.~{Hu} and D.~{Huterer},
\newblock Phys. Rev. {\bf D79}, 023004 (2009), [\eprintmod[0810.1744]],
\newblock (MHH1).

\bibitem{PaperII}
M.~J. Mortonson, W.~Hu and D.~Huterer,
\newblock Phys. Rev. {\bf D81}, 063007 (2010), [\eprintmod[0912.3816]],
\newblock (MHH2).
%%CITATION = 0912.3816;%%

\bibitem{Huterer_Peiris}
D.~Huterer and H.~V. Peiris,
\newblock Phys. Rev. {\bf D75}, 083503 (2007), [\eprintmod[astro-ph/0610427]].
%%CITATION = ASTRO-PH/0610427;%%

\bibitem{Tang:2008hm}
J.~Tang, F.~B. Abdalla and J.~Weller,
\newblock \eprintmod[0807.3140].
%%CITATION = 0807.3140;%%

\bibitem{Kitching:2009yr}
T.~D. Kitching and A.~Amara,
\newblock \eprintmod[0905.3383].
%%CITATION = 0905.3383;%%

\bibitem{dePutter_Linder2}
R.~de~Putter and E.~V. Linder,
\newblock \eprintmod[0812.1794].
%%CITATION = 0812.1794;%%

\bibitem{Barnard:2008mn}
M.~Barnard, A.~Abrahamse, A.~J. Albrecht, B.~Bozek and M.~Yashar,
\newblock Phys. Rev. {\bf D78}, 043528 (2008), [\eprintmod[0804.0413]].
%%CITATION = 0804.0413;%%

\bibitem{SCP_Union}
M.~Kowalski {\em et~al.},
\newblock Astrophys. J. {\bf 686}, 749 (2008), [\eprintmod[0804.4142]].
%%CITATION = 0804.4142;%%

\bibitem{Union_like}
\url{http://supernova.lbl.gov/Union/}.

\bibitem{Komatsu_2008}
E.~Komatsu {\em et~al.},
\newblock Astrophys. J. Suppl. {\bf 180}, 330 (2009), [\eprintmod[0803.0547]].
%%CITATION = 0803.0547;%%

\bibitem{Nolta_2008}
M.~R. Nolta {\em et~al.},
\newblock Astrophys. J. Suppl. {\bf 180}, 296 (2009), [\eprintmod[0803.0593]].
%%CITATION = 0803.0593;%%

\bibitem{Dunkley_2008}
J.~Dunkley {\em et~al.},
\newblock Astrophys. J. Suppl. {\bf 180}, 306 (2009), [\eprintmod[0803.0586]].
%%CITATION = 0803.0586;%%

\bibitem{WMAP_like}
\url{http://lambda.gsfc.nasa.gov/}.

\bibitem{Lewis:1999bs}
A.~Lewis, A.~Challinor and A.~Lasenby,
\newblock Astrophys. J. {\bf 538}, 473 (2000), [\eprintmod[astro-ph/9911177]].
%%CITATION = ASTRO-PH/9911177;%%

\bibitem{camb_url}
\url{http://camb.info/}.

\bibitem{PPF}
W.~{Fang}, W.~{Hu} and A.~{Lewis},
\newblock Phys. Rev. {\bf D78}, 087303 (2008), [\eprintmod[0808.3125]].

\bibitem{ppf_url}
\url{http://camb.info/ppf/}.

\bibitem{KLMM_SNAP}
A.~G. {Kim}, E.~V. {Linder}, R.~{Miquel} and N.~{Mostek},
\newblock \mnras {\bf 347}, 909 (2004), [\eprintmod[astro-ph/0304509]].

\bibitem{SNAP}
G.~Aldering {\em et~al.},
\newblock \eprintmod[astro-ph/0405232].
%%CITATION = ASTRO-PH 0405232;%%

\bibitem{LinHut_highz}
E.~V. Linder and D.~Huterer,
\newblock Phys. Rev. {\bf D67}, 081303 (2003), [\eprintmod[astro-ph/0208138]].
%%CITATION = ASTRO-PH/0208138;%%

\bibitem{dePutter:2009kn}
R.~de~Putter, O.~Zahn and E.~V. Linder,
\newblock Phys. Rev. {\bf D79}, 065033 (2009), [\eprintmod[0901.0916]].
%%CITATION = 0901.0916;%%

\bibitem{Hollenstein:2009ph}
L.~Hollenstein, D.~Sapone, R.~Crittenden and B.~M. Schaefer,
\newblock JCAP {\bf 0904}, 012 (2009), [\eprintmod[0902.1494]].
%%CITATION = 0902.1494;%%

\bibitem{Doran}
M.~{Doran}, G.~{Robbers} and C.~{Wetterich},
\newblock Phys. Rev. {\bf D75}, 023003 (2007), [\eprintmod[astro-ph/0609814]].

\bibitem{Eisenstein}
D.~J. Eisenstein {\em et~al.},
\newblock Astrophys. J. {\bf 633}, 560 (2005), [\eprintmod[astro-ph/0501171]].
%%CITATION = ASTRO-PH 0501171;%%

\bibitem{Percival09}
W.~J. Percival {\em et~al.},
\newblock Mon. Not. Roy. Astron. Soc. {\bf 401}, 2148 (2010),
  [\eprintmod[0907.1660]].
%%CITATION = 0907.1660;%%

\bibitem{SHOES}
A.~G. Riess {\em et~al.},
\newblock Astrophys. J. {\bf 699}, 539 (2009), [\eprintmod[0905.0695]].
%%CITATION = 0905.0695;%%

\bibitem{HubTrans}
M.~Mortonson, W.~Hu and D.~Huterer,
\newblock Phys. Rev. {\bf D80}, 067301 (2009), [\eprintmod[0908.1408]].
%%CITATION = 0908.1408;%%

\bibitem{Christensen:2001gj}
N.~Christensen, R.~Meyer, L.~Knox and B.~Luey,
\newblock Class. Quant. Grav. {\bf 18}, 2677 (2001),
  [\eprintmod[astro-ph/0103134]].
%%CITATION = ASTRO-PH 0103134;%%

\bibitem{Kosowsky:2002zt}
A.~Kosowsky, M.~Milosavljevic and R.~Jimenez,
\newblock Phys. Rev. {\bf D66}, 063007 (2002), [\eprintmod[astro-ph/0206014]].
%%CITATION = ASTRO-PH 0206014;%%

\bibitem{Dunetal05}
J.~{Dunkley}, M.~{Bucher}, P.~G. {Ferreira}, K.~{Moodley} and C.~{Skordis},
\newblock \mnras {\bf 356}, 925 (2005), [\eprintmod[astro-ph/0405462]].

\bibitem{gelman/rubin}
A.~Gelman and D.~Rubin,
\newblock Statistical Science {\bf 7}, 452 (1992).

\bibitem{Lewis:2002ah}
A.~Lewis and S.~Bridle,
\newblock Phys. Rev. {\bf D66}, 103511 (2002), [\eprintmod[astro-ph/0205436]].
%%CITATION = ASTRO-PH 0205436;%%

\bibitem{cosmomc_url}
\url{http://cosmologist.info/cosmomc/}.

\bibitem{Linder_wa}
E.~V. Linder,
\newblock Phys. Rev. Lett. {\bf 90}, 091301 (2003),
  [\eprintmod[astro-ph/0208512]].
%%CITATION = ASTRO-PH 0208512;%%

\bibitem{Chevallier_Polarski}
M.~Chevallier and D.~Polarski,
\newblock Int. J. Mod. Phys. {\bf D10}, 213 (2001),
  [\eprintmod[gr-qc/0009008]].
%%CITATION = GR-QC 0009008;%%

\bibitem{Bassett:2009uv}
B.~A. Bassett, Y.~Fantaye, R.~Hlozek and J.~Kotze,
\newblock \eprintmod[0906.0993].
%%CITATION = 0906.0993;%%

\bibitem{Davetal07}
T.~M. {Davis} {\em et~al.},
\newblock \apj {\bf 666}, 716 (2007), [\eprintmod[astro-ph/0701510]].

\bibitem{Wright2007}
E.~L. Wright,
\newblock Astrophys. J. {\bf 664}, 633 (2007), [\eprintmod[astro-ph/0701584]].
%%CITATION = ASTRO-PH/0701584;%%

\bibitem{Mantz}
A.~Mantz, S.~W. Allen, D.~Rapetti and H.~Ebeling,
\newblock \eprintmod[0909.3098].
%%CITATION = 0909.3098;%%

\bibitem{Caldwell_Linder}
R.~R. Caldwell and E.~V. Linder,
\newblock Phys. Rev. Lett. {\bf 95}, 141301 (2005),
  [\eprintmod[astro-ph/0505494]].
%%CITATION = ASTRO-PH 0505494;%%

\bibitem{AlbSko00}
A.~{Albrecht} and C.~{Skordis},
\newblock Phys. Rev. Lett. {\bf 84}, 2076 (2000),
  [\eprintmod[astro-ph/9908085]].

\end{thebibliography}

\end{document}